\documentclass[preprint,5p,twocolumn]{elsarticle}
\usepackage{hyperref}
\usepackage[linesnumbered, ruled]{algorithm2e}
\usepackage{algpseudocode}
\usepackage{amsmath}
\usepackage{color}
\usepackage{setspace}

\usepackage{url}
\usepackage{booktabs}
\usepackage{lscape}
\usepackage{multirow}
\usepackage[normalem]{ulem}
\usepackage{geometry}
\usepackage{setspace}
\usepackage{amsmath,amsthm,txfonts}
\usepackage{caption}
\usepackage{subfigure}
\usepackage{comment}
\useunder{\uline}{\ul}{}
\theoremstyle{definition}
\newtheorem{definition}{Definition}[section]

\journal{Information and Software Technology}
\begin{document}

\begin{frontmatter}

\title{Automatically detecting the conflicts between software requirements \\ based on finer semantic analysis}

\author[address]{Weize Guo}
\ead{guoweize@buaa.edu.cn}

\author[address]{Li Zhang}
\ead{lily@buaa.edu.cn}

\author[address]{Xiaoli Lian\corref{corresponding_author}}
\cortext[corresponding_author]{Corresponding author}
\ead{lianxiaoli@buaa.edu.cn}

\address[address]{Software Engineering Institute, Beihang University, Beijing China}

\begin{abstract}
    \textbf{Context:}
    Conflicts between software requirements bring uncertainties to product development.
    Some great approaches have been proposed to identify these conflicts.
    However, they usually require the software requirements represented with specific templates and/or depend on other external source which is often uneasy to build for lots of projects in practice.
    
    \noindent \textbf{Objective:}
    We aim to propose an approach \underline{F}iner \underline{S}emantic \underline{A}nalysis-based \underline{R}equirements \underline{C}onflict Detector (\emph{FSARC}) to automatically detecting the conflicts between the given natural language functional requirements by analyzing their finer semantic compositions. 
    
    \noindent \textbf{Method:}
    Firstly, we build a harmonized semantic meta-model of functional requirements with the form of eight-tuple.
    Then we propose algorithms to automatically analyze the linguistic features of requirements and to annotate the semantic elements for their semantic model construction.
    And we define seven types of conflicts as long as their heuristic detecting rules on the ground of their text pattern and semantical dependency.
    Finally, we design and implement the algorithm for conflicts detection.
    
    \noindent \textbf{Results:}
    The experiment with three different open requirements datasets shows that our semantic analyzing algorithms can correctly identify about 94.93\% elements of software requirements in average.
    And the other experiment with four requirement datasets illustrates that the recall of \emph{FSARC} is nearly 100\% and the average precision is 83.88\% on conflicts detection.
    
    \noindent \textbf{Conclusion:}
    We provide a useful tool for detecting the conflicts between natural language functional requirements to improve the quality of the final requirements set.
    Besides, our approach is capable of transforming the natural language functional requirements into eight semantic tuples, which is useful not only the detection of the conflicts between requirements but also some other tasks such as constructing the association between requirements and so on. 
\end{abstract}

\begin{keyword}
    Software requirement specifications \sep Requirements conflict \sep Semantic requirement metamodel \sep Semantic elements labeling
\end{keyword}

\end{frontmatter}

\section{Introduction}

Software requirements, especially of the large and complex projects, are usually from massive stakeholders with diverse background and interests \cite{Pohl2010Requirements}.
Conflicts among software requirements are therefore inevitable due to their different views and concerns.
Besides, requirements are continually changing \cite{Jayatilleke2018systematic}.
Conflicts occur during the whole life-cycle of software product. Actually, some researches show that a very high number of conflicting requirements was identified among software projects in practice \cite{Schneider1992experimental, Ramzan2010Intelligent}.
Egyed and Grünbacher \cite{Egyed2004Identifying} pointed out that ``requirements conflict with each other if they make contradicting statements about common software attributes \dots Given that there may be up to $n^2$ conflicts among $n$ requirements \dots, the number of potential conflicts, \dots, could be enormous, burdening the engineer with the time-intensive and error-prone task of identifying the true conflicts''.
And Mairiza et al. \cite{Mairiza2010ontological} reported that the ratio of conflicting requirements is from 40\% to 60\%.
The conflicts between requirements cause the project cost overrun \cite{Yang2008Survey, Urbieta2011Detecting} and failure \cite{Butt2011Requirement}.
However, detecting the conflicts of requirements is one of the most important but yet one of the most challenging aspects of requirements validation \cite{Davis1990Software, Pohl1993three, Zowghi2002three}.

The current methods of detecting requirements conflicts can be divided into two categories: manual walk-through and automatic detection.
The first stream of manual approaches, such as the win-win requirements negotiation approach \cite{Boehm1996Identifying} and its variant easy-win-win \cite{Briggs2002EasyWinWin}, are very common in practice.
However, many experts with enough domain knowledge are usually involved and need to spend tedious time on it. The manual identification is therefore ``time-intensive and error-prone'' \cite{Egyed2004Identifying}.
Relatively speaking, the automated genre can save human efforts to a certain degree \cite{Egyed2004Identifying}.
However, most of these approaches need the requirements with specific format such as the extended Bakkus-Naur-Form (EBNF)\cite{Moser2011Requirements} and the Semantic Application Design  Language (SADL) \cite{Moitra2018Towards}.
Besides, some other artifacts beyond requirement specifications are required such as executable code \cite{Egyed2004Identifying}, system performance constraints \cite{Walter2017formalization} which are usually unavailable  especially in the early phrase of projects.

In addition, the conflicts under analysis in these research are different, including redundant requirements \cite{Walter2017formalization}, conflicting or incomplete requirements \cite{Moitra2018Towards} and so on.
And it is necessary to build an harmonized view about the conflict types.

The final goal of this study is to automatically detect the conflicts between natural language functional requirements.
We regard natural language as the requirement format of our study, since 71.8\% of the requirement documents are written in natural language in engineering practice\cite{Luisa2004Market}.
Natural language requirements may have variety of forms, including modal verb-style statements (such as \emph{shall-} or \emph{will-} style), use cases, user stories and feature lists \cite{Pohl2015Requirements}.
And \emph{shall}-style statements are very commonly used to specify requirements in industry \cite{Whittle2009RELAX}.
Besides, we mainly concern ourselves with \emph{functional requirements} which may have their related constraints in resource limitation, access time and so on, since they are the base of software design and development.
And the constraint from non-functional requirements need to be satisfied through the implementation of functional requirements.
So in the remainder of this study, requirements mean a set of NL \emph{shall}-style statements on the functionality of the system or system components.

In order to achieve our goal, we propose an approach of a \underline{F}iner \underline{S}emantic \underline{A}nalysis-based \underline{R}equirements \underline{C}onflict Detector (\emph{FSARC}) to identify seven types of conflicts among functional natural language requirements based on analyzing and detecting their finer semantical components.
To be specific, we firstly build a semantic meta-model of software requirements denoting the semantic elements of functional requirements.
Then we design heuristic rules and algorithms for automatically identifying these elements based on the natural language processing results such as part-of-speech (POS) and the semantic dependency parsing (SDP).
Finally, we define seven types of conflicts on the ground of analyzing the existing research and detect them based on a set of predefined rules.

Our contributions are therefore threefold:

\begin{enumerate}[(1)]
    \item
    We provide a harmonized view of the semantic meta-model of software functional requirements, and propose an approach of automatically identifying the related elements from NL requirements.
    \item
    We define seven types of conflicts among requirements with three categories through carefully analyzing the related literatures, and propose an approach of automatically detecting them.
    \item
    The two automated approaches for semantic elements identification and conflicts detection respectively were evaluated with more than three open requirement sets.
    And the results show the great performance of our approaches. We public the source code of these two algorithms via Github \url{https://github.com/GuoWeize/FSARC}.
\end{enumerate}

We organize the structure of this present paper as follows.
Section \ref{sec:backgroundRelatedWorks} introduces the background on the general definition of requirements conflict and the related work on the detection of requirements conflict.
Section \ref{sec:approachFramework} describes the framework of our approach \emph{FASRC}.
Then Section \ref{sec:structuring} and \ref{sec:conflictDetection} illustrate the two parts of \emph{FASRC} including the semantic elements detection of software requirements and the conflicts detection between them. In the end of each section we introduce the related evaluation experiments and the results. In Section \ref{sec:threats} we discuss the threats to validity of this study. In the end, the conclusions of this work and the future work plan are given in Section \ref{sec:conclusion}.

\section{Background and the related works}
\label{sec:backgroundRelatedWorks}

\subsection{The general definition of requirements conflicts}

In the field of requirement engineering, the term \emph{conflict} involves inference, interdependency, and inconsistency between requirements \cite{Mairiza2009Managing}.
Kim et al. \cite{Kim2007Managing} regarded requirements conflicts as ``the interactions and dependencies between requirements that can lead to negative or undesired operation of the system''.
Alberto et al. \cite{Sardinha2010Conflict} believed that ``when a crosscutting requirement has a negative contribution with another crosscutting requirement in the same base requirement, a conflict between aspect-oriented software requirements occurs''.
Cameron et al. \cite{Cameron1993Feature} regarded requirements conflicts as ``unexpected or contradictory interaction between requirements that has a negative effect on the results''.

From the perspective of conflicting objects, the conflicts may exist between the requirements of the same or different types, such as between functional requirements \cite{hausmann2002detection}, between functional and non-functional requirements \cite{Egyed2004Identifying}, and between non-functional requirements \cite{Sadana2007Analysis}.
Also conflicts can exist between  two different manifestations of the same requirement, such as between textual term and use case graph \cite{Kamalrudin2010Managing}, between textual term and table form \cite{Salado2015Tension}, and so on.

In this study, we focus on the semantical conflicts among different textual functional requirements, which means that \emph{the inference, interdependency and any other relationship between requirements that can lead to inconsistency.}
For instance, the duplication of two requirements may cause inconsistency when one requirement is changed  while another not.

\subsection{The NLP technologies used in this study}

The natural language processing (NLP) technologies are used to aid analyzing and understanding the human language.
It involves lexical and syntactic analysis to obtain the meaning of the strings of symbols in natural language conforming to the rules of formal grammar.
In this study, we would like to analyze the structure and meaning of textual functional requirement statement, helping detect conflicts further. With the help of NLP, we are able to identify the semantic elements of requirement statement through analyzing their grammar and syntactic structures, and to build the mapping rules between these structure units and the semantic elements.

We mainly use Part-Of-Speech (POS) tagging and Dependency Parsing (DS), especially Semantic Dependency Parsing (SDP).
POS is to mark each word with a correct part of speech, that is, to determine whether the word is a noun, a verb, an adjective or others.
SDP is to analyze the binary semantic relationship between the language units (i.e., word or phrase) of one sentence, for instance, a \emph{nsubj} (subject to predicate) relationship existing from ``like'' to ``I'' in the sentence ``I like it.''.

CoreNLP \cite{CoreNLP2014}, proposed by Stanford University, is one of the most popular NLP tools, providing many kinds of basic analysis for several languages, including POS tagging, parsing tree analysis, dependency relationship analysis and so on.
And most of its POS and SDP analysis results are followed the definitions of Universal Dependencies \footnote{\url{https://universaldependencies.org/}}.
Basic POS tags include N(noun), V(verb), ADJ(adjective), ADV(adverb) and so on.
Basic dependencies in SDP include \emph{nsubj} (subject to predicate), \emph{dobj} (predicate to direct object), \emph{root} (root to predicate) and so on.
And other dependencies are mainly about the relation between modifier, qualifier and clause.
We firstly pre-process textual requirements by CoreNLP, and then combine the results of POS and SDP to obtain a comprehensive view of their semantic roles.

\subsection{Related works on requirements conflict detection}

Detecting the conflicts between software requirements has always been a hot topic in software engineering, and there have been a few pretty good studies.

Egyed and Grünbacher \cite{Egyed2004Identifying} detected the cooperating and conflicting relations between natural language requirements based on the trace dependencies among requirements, with the usage of a tool-supported trace analysis technique \cite{egyed2003scenario}.
Besides the NL requirements, they also need the associations between requirements and test scenarios as the input.

Kim et al. \cite{Kim2007Managing} require requirements described in goals and scenarios using the authoring structure, which contains Action (Verb) + Object (Object) + Resource (Resource).
They can automatically detect two types of conflicts which are syntactic conflict meaning requirements conflicts according to a predefined condition, such as ``Different Verb $\bigcap$ Same object'', ``Same Verb $\bigcap$ Different Object'', or ``Same Resource'', and semantic conflict including activity conflicts and resource conflicts.

Moser et al. \cite{Moser2011Requirements} proposed an ontology-based approach OntReq to detect three types of conflicts: conflict between requirements (CRR), conflict between a requirement with a constraint (CRC), conflict between a requirement with a formal guideline(CRG), i.e., ill-formed requirement.
They require functional requirements and constraints, including technical constraints, requirement constraints, documentation guidelines and glossary as inputs.
And functional requirements must follow the EBNF template, specifying under which conditions a target system should provide a certain functionality regarding a specific object to a certain entity or role.

Ali et al. \cite{Ali2013Reasoning} applied SAT-based techniques \cite{biere2009handbook} to automatically check the consistency between contextual requirements and the consistency between executable tasks.
They require requirements represented by a contextual goal model, which is the environment in which the system operates.
And they defined two conflict types: conflicting changes (many processes try simultaneously to change the same object into different states) and exclusive possession (processes need an exclusive possession of an object).

Chentouf \cite{Chentouf2014Managing} require requirements in a controlled natural language derived from KAOS \cite{dardenne1993goal}.
He focused on three conflict types: duplicated requirement (two requirements being exactly the same or one being included in the other), incompatible requirements (two requirements are ambiguous, incompatible, or contradictory), assumption alteration (the output of one requirement’s operation is part of the inputs or outputs of the other’s operation).
Unfortunately, this work doesn't include an automated approach or the critical detection algorithms.

Salado and Nilchiani \cite{Salado2015Tension}  traced requirements to stakeholders' needs, and automatically constructed the tension matrix (including resources, phases of matter, laws of physics, laws of society, and logical contradictions) to identify potential conflicting requirements.
Based on the tension matrix, they defined five conflicts types from resources, phases of matter, laws of physics, laws of society and logical contradictions.
Their method is particularly effective for the requirements containing quantitative numbers.

Walter et al. \cite{Walter2017formalization} formalized NL requirements into conjunctive normal form (CNF) using specification pattern systems (SPS), linear temporal logic (LTL) and first order logic (FOL) semi-automatically.
They only focus on redundant requirement specifications.

Moitra et al. \cite{Moitra2018Towards} need the requirements specified with Semantic Application Design Language (SADL) \cite{crapo2013toward}.
They perform formal analysis using ACL2 Sedan tool \cite{chamarthi2011acl2} automatically and resolve errors during the requirements authoring process. According to their formal rules, they only detect conflicting or incomplete requirements. 

From the above simple survey, we can make the following observations:
\begin{itemize}
    \item
    The \emph{input formats of requirements} can be mainly divided into two categories: in natural language and in specific format, usually constraint format.
    Given that most software requirements are in natural language in practice \cite{Luisa2004Market}, the approaches with specific formal format need more transforming work for practical purposes.
    However, the transforming work is often unavailable.
    \item
    Many methods require \emph{external data} to assist or promote requirement formalization and inconsistence detection, such as the tracing between requirements and test scenarios \cite{egyed2003scenario}, glossary \cite{Moser2011Requirements} and stakeholder' needs \cite{Salado2015Tension}.
    However, some data is usually hard to acquire, even unavailable \cite{Arora2019Active}.
    \item
    These work focus on \emph{different kinds of conflicts} from diverse perspectives, although with small portion of overlapping.
    \item
    Only \cite{Chentouf2014Managing} and \cite{Salado2015Tension} provide \emph{public dataset}. Unavailable data causes inconvenience of the work reproducing and testing.
\end{itemize}

\section{FASRC: Finer Semantic Analysis-based Requirements Conflicts Detector}
\label{sec:approachFramework}

\begin{figure*}[htb]
	\centering
	\includegraphics[trim = {0cm 3cm 0cm 0cm}, clip, width=0.95\textwidth]{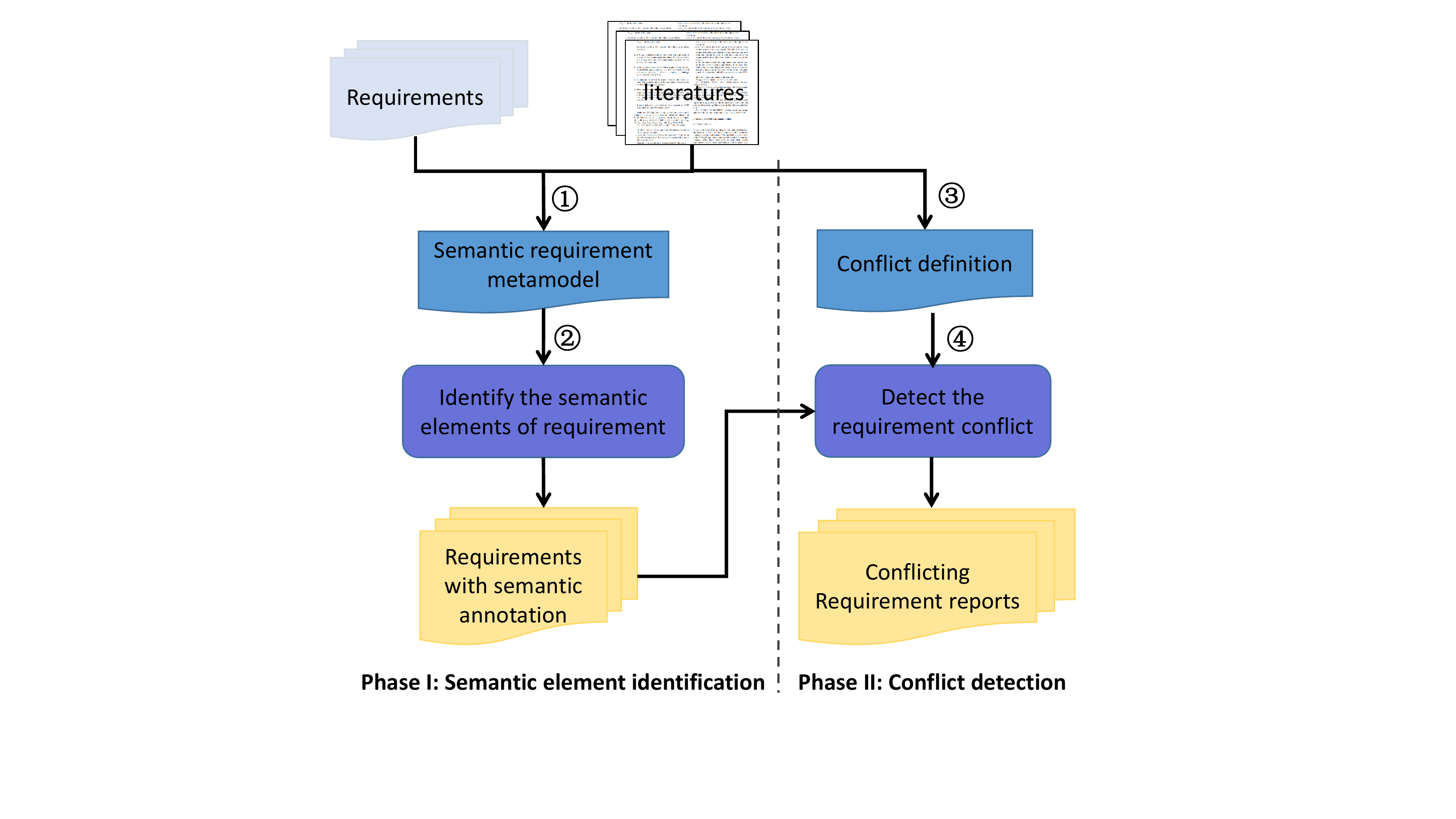}
	\caption{Our approach for conflict detection}
	\label{fig:framework}
\end{figure*}

Our approach \emph{FASRC} for requirements conflict detection involves two phases, which are semantic element identification (Phase I) and conflicts detection (Phase II).
The procedures of our approach is shown in \textbf{Figure} \ref{fig:framework}.

In the first phase, we identify the finer semantic elements of NL requirements.
For this purpose, the semantic requirement meta-model defining the common compositions of the semantic elements is required.
Thus, we firstly provide an harmonized view of the semantic requirement meta-model through analyzing the existing literatures and part of the requirement set (in Step \textcircled{1}).
Then, we propose an algorithm for identifying the semantic elements for each requirement (in Step \textcircled{2}).
In other words, we annotate the requirements with their semantic elements automatically.
The core step is to discover and build the mapping rules between the grammar and semantic elements of requirements.
To implement this, we manually annotated the semantic elements of a random requirement set of Unmanned Aerial Vehicle (UAV) from the University of Notre Dame \cite{Cleland2018Dronology}, and looked up their grammar elements on the ground of POS and SDP generated by Stanford CoreNLP \cite{CoreNLP2014}.
We analyzed the mapping between the manual annotations and the NLP results, and built the rules, which are the key part of our semantic annotating algorithm.

In the second phase, we detect the conflicts between requirements based on their finer semantic elements annotation.
Obviously, the formal definition of our focused conflicts is the prerequisite for the conflict detection.
We define seven conflict types with three categories in this study (in Step \textcircled{3}).
Finally, we propose our conflict detection algorithm by defining the rules upon the conflict definition and the semantic annotations of requirements (in Step \textcircled{4}).

\section{Phase I: Automatically identify the finer semantic elements of NL requirements}
\label{sec:structuring}

\subsection{Our semantic metamodel of functional requirements}
\label{subsec:semanticDefinition}
There are some existing semantic models of requirements. 
Kim et al. \cite{Kim2007Managing} proposed that requirement statements can be described with goals and scenarios authoring structure, which contains Action (Verb) + Object + Resource.
Rupp et al. \cite{Rupp2009Requirements} proposed an EBNF template of requirements specifying under which conditions (under condition) a target system should provide a certain functionality (process) regarding a specific object (thing to be processed) to a certain entity or role (somebody or something).
Chentouf \cite{Chentouf2014Managing} proposed that requirement statements can be formally described as eight tuples including the ID, the effect of requirement (i.e., start, stop, or forbid an operation), the trigger, agent, operation, input, output, and the waiting time for the periodical execution of operation.

We prefer Chentouf's model because they give finer elements such as the input, output and the execution frequency.
However, there are three points which can be improved:

\begin{itemize}
    \item
    Although the effect of requirements (i.e., \emph{start}, \emph{stop} or \emph{forbid} an operation) can explicitly tell whether two requirements are conflict or not on the state, they didn't give the way of specifying this effect. Besides, it is redundant to the \emph{operation} semantically.
    \item
    They only describe the operation frequency of the operation.
    However, there are much more restriction about the operation such as the execution sequence and resource limitation.
    \item
    They didn't give formal definition of these eight tuples but in natural language only.
    This brings ambiguity.
    For example, the input is constituted by the set of objects, but the definition of object is unmentioned.
\end{itemize}

We therefore present our novel eight-tuple: \{\emph{id, groupId, event, agent, operation, input, output, restriction}\}.
The specific definition is as follows:

\begin{itemize}
    \item[--]\texttt{id:}
    The identifier of the requirement, usually consisting of numbers and/or letters.
    The \texttt{id} of each requirement should be unique.
    \item[--]\texttt{groupId:}
    For the sake of simplicity, we expect one requirement contains only one operation.
    Thus, we reorganize the requirements with multiple predicate and split them into several associated requirements of the same group with the same \texttt{groupID}, \texttt{agent} and \texttt{event}.
    The \texttt{groupId} is a natural number.
    \item[--]\texttt{event:}
    The timing or trigger condition of one requirement.
    When \texttt{event} is satisfied, \texttt{agent} must perform the \texttt{operation}.
    We find that \texttt{event} is usually a complete sentence (i.e., adverbial clause), like the requirement statement, but without \texttt{id}, \texttt{groupId} and \texttt{event}.
    Therefore, we use five-tuple: \{\texttt{agent}, \texttt{operation}, \texttt{input}, \texttt{output}, \texttt{restriction}\} to describe the semantic elements of \texttt{event}.
    If there are multiple \texttt{conditions} in one single requirement, five-tuple of all \texttt{events} are connected by ``\emph{and}'' or ``\emph{or}'', which depends on the conjunction between two \texttt{event}s.
    \item[--]\texttt{agent:}
    The executor of \texttt{operation}, usually the real subject of the main clause of the requirement.
    \item[--]\texttt{operation:}
    The core action of the requirement, also the predicate of the main clause of the requirement.
    The action is usually depicted with a verb and sometimes contains other supplements like \emph{to do}.
    \item[--]\texttt{input:}
    All the data that already exists before, which needs to be used when the \texttt{operation} is executed.
    \item[--]\texttt{output:}
    All objects that can be created, destroyed, or altered after executing the \texttt{operation}.
    \item[--]\texttt{restriction:}
    Constraint is on performing the \texttt{operation}, which can be the execution time, place, frequency, execution sequence, and some other restriction such as resource and its quantity.
\end{itemize}

\begin{figure}[htb]
    \centering
    \includegraphics[trim = {1cm 5cm 0cm 4cm}, clip, width=0.45\textwidth]{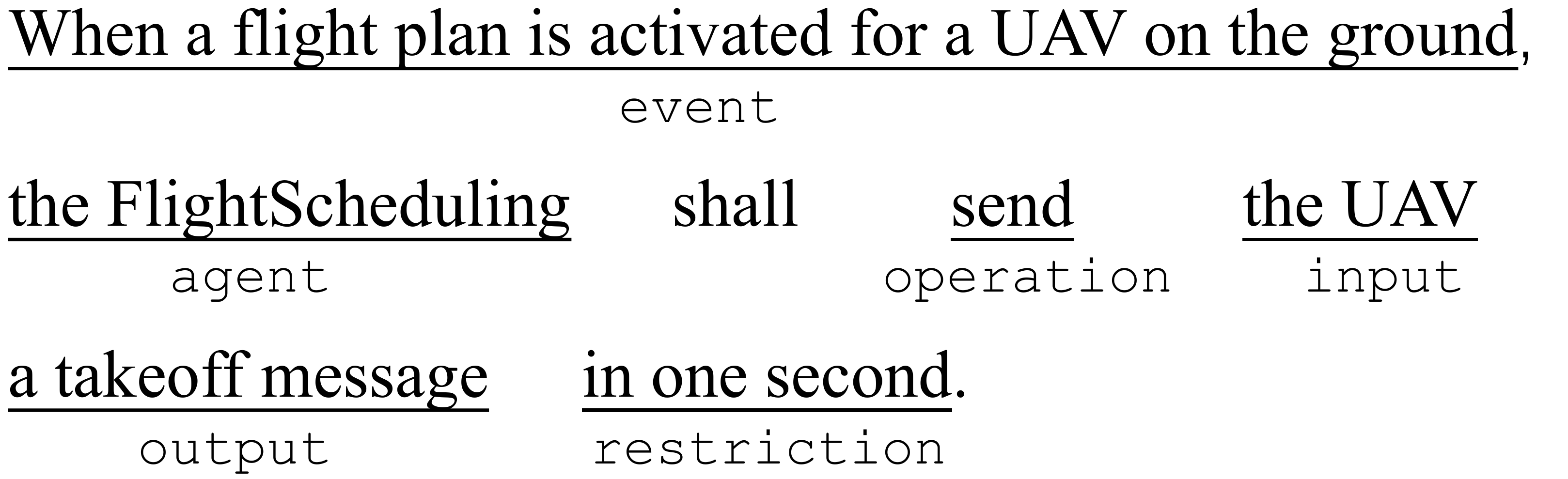}
    \caption{The eight-tuple of one requirement example in UAV set}
    \label{fig:eighttuplesexample}
\end{figure}

To make it easy to understand, we give an example of the eight-tuple of a requirement of UAV in \textbf{Figure} \ref{fig:eighttuplesexample}.
In this example, we annotate the \texttt{event}, \texttt{agent}, \texttt{operation}, \texttt{input}, \texttt{output} and \texttt{restriction}.
The \texttt{id} and \texttt{groupId} rely on other requirements in the whole set and we don't list them here.

We find that \texttt{agent} is a nounal structure, usually a noun. Sometimes there are some modifiers to give the indispensable supplement of the nounal structure.
The modifiers may include adjective word, adjective clause, determiner, quantifier and so on.
Also, the subordination between this and another nounal structure is  included in the modifiers, since it specifies the nounal structure too.
And \texttt{input} and \texttt{output} are both consisted of a series of nounal structures.
For convenience of formalization, we define a structure called \texttt{entity}, which denoted the nounal structure above.
Therefore, \texttt{agent} is an \texttt{entity}, \texttt{input} and \texttt{output} are sets of \texttt{entities}.

To facilitate automatically identifying the semantic tuples, we present the definition of eight-tuple in the form of BNF as follows:

\small
\noindent\texttt{<requirement>} \textbf{::=} \texttt{<id>} \texttt{<groupId>} \texttt{<event>} \texttt{<agent>} \texttt{<operation>} \texttt{<input>} \texttt{<output>} \texttt{<restriction>}
\\\texttt{<id>} \textbf{::=} \texttt{<letter>}\texttt{<string>}
\\\texttt{<groupId>} \textbf{::=} \texttt{<digit>}\{\texttt{<digit>}\}
\\\texttt{<event>} \textbf{::=} $\varnothing$ $\vert$ \texttt{<condition>} \{\texttt{<conj> <condition>}\}
\\\texttt{<agent>} \textbf{::=} $\varnothing$ $\vert$ \texttt{<entity>}
\\\texttt{<operation>} \textbf{::=} \texttt{<operation\_mode>} \texttt{<predicate>}
\\\texttt{<input>} \textbf{::=} \texttt{<entity\_set>}
\\\texttt{<output>} \textbf{::=} \texttt{<entity\_set>}
\\\texttt{<restriction>} \textbf{::=} $\varnothing$ $\vert$ \{\texttt{<string>}\}$^+$
\\\texttt{<letter>} \textbf{::=} ``\emph{A}'' $\vert$ \dots $\vert$ ``\emph{Z}'' $\vert$ ``\emph{a}'' $\vert$ \dots $\vert$ ``\emph{z}''
\\\texttt{<digit>} \textbf{::=} ``\emph{0}'' $\vert$ \dots $\vert$ ``\emph{9}''
\\\texttt{<string>} \textbf{::=} \{ \{\texttt{<digit>}\}\{\texttt{<letter>}\} \}
\\\texttt{<condition>} \textbf{::=} \texttt{<agent>} \texttt{<operation>} \texttt{<input>} \texttt{<output>} \texttt{<restriction>}
\\\texttt{<conj>} \textbf{::=} ``\emph{and}'' $\vert$ ``\emph{or}''
\\\texttt{<entity>} \textbf{::=}  \texttt{<modifier>} \texttt{<base>}
\\\texttt{<modifier>} \textbf{::=} $\varnothing$ $\vert$ $\{\texttt{<string>}\}^+$
\\\texttt{<base>} \textbf{::=} \texttt{<string>}
\\\texttt{<operation\_mode>} \textbf{::=} $\varnothing$ $\vert$ \emph{ABLE} $\vert$ \emph{NOT}
\\\texttt{<predicate>} \textbf{::=} \texttt{<string>}
\\\texttt{<entity\_set>} \textbf{::=} $\varnothing$ $\vert$ $\{\texttt{<entity>}\}^+$
\normalsize

According to the above BNF items, one \texttt{requirement} is an eight-tuple.
\texttt{Event} is usually a single clause depicting the \texttt{condition}, but if there are multiple \texttt{conditions} in a single requirement, they should be connected by \texttt{conj} (``\emph{and}'' or ``\emph{or}'') based on the conjunction between them.
And \texttt{event} is labeled as ``\emph{ALL}'' when no actual \texttt{event} can be found in a requirement, which means the \texttt{operation} can be executed under any circumstance.
Given that conflict detection upon clauses are not easy since clause is usually composed of a couple of semantic units, we represent the \texttt{condition} as five-tuple. \texttt{Operation} is a string of core verb with a specific \texttt{operation\_mode}: ``\emph{ABLE}'', ``\emph{NOT}'' or the default mode, and we will specify them in Section \ref{subsubsec:opeIdentify}.
The \texttt{agent},\texttt{input} and \texttt{output} are composed of noun words/phrases and their modifiers. For the convenience of describing these tuples, we refer the noun word/phrase (base) and all of its modifiers as \texttt{entity}.
\texttt{Restriction} is defined as a set of \texttt{string} and each \texttt{string} denotes a single constraint.

\subsection{Automatically identifying the finer semantic elements}
\label{subsec:semanticElementIden}

In general, our algorithm is based on heuristic rules and we adopt the divide and conquer principle during the algorithm design.
Since each semantic element is distinctive, we propose different algorithms to identify them.
The pseudo-code description of our main algorithm is given below in \textbf{Alg.} \ref{alg:mainAlg}.

\begin{algorithm}[htb]
	\caption{MainAlgorithm}
	\label{alg:mainAlg}
	\KwIn {requirement (one NL requirement entry), id, groupId}
	\KwOut {requirement\_tuple (eight-tuple)}
	\tcc{Use POS and SDP to parse requirement statement}
	(pos, dependencies) = \emph{CoreNLPParse}(requirement)\\
	\If {\rm the requirement has multiple predicate verbs} {
		\tcc{Divide it into multiple requirements}
		sentence\_group = \emph{SentenceSplit}(requirement)\\
		groupId = groupId + 1\\
		\tcc{Modelling each divided requirement}
		\For {\rm \textbf{each} sentence $\in$ sentence\_group} {
			r = \emph{MainAlgorithm}(sentence, id, groupId)\\
			\textbf{return} r
		}
	}
	\tcc{Extract each tuple from the clause}
	operation = \emph{OperationParse} (requirement, pos, dependencies)\\
	agent = \emph{AgentParse}(operation, dependencies)\\
	event = \emph{EventParse}(requirement, agent)\\
	(input, output) = \emph{InputOutputParse}(operation, dependencies)\\
	restriction = \emph{RestrictionParse}(requirement, pos, dependencies)\\
	requirement\_tuple = \{id, groupId, event, agent, operation, input, output, restriction\}\\
	\textbf{return} requirement\_tuple
\end{algorithm}

\textbf{Heuristic rules construction}  
For identifying each of the elements, except the \texttt{id}, we need to construct the extraction rule firstly.
To be specific, we randomly selected 50 requirements from the UAV set.
The first two authors manually annotated their semantic elements independently and discussed the discrepancies to form an agreement on the final annotations.
Then we run CoreNLP\cite{CoreNLP2014} and generated POS as well as SDP results of these requirements.
We recorded and analyzed the mapping between each semantic elements of each annotated requirements and their grammar elements.
Finally we gave the heuristic rules of identifying each semantic element from the NL requirements.

During the analysis, we have three findings about the NL requirements, which can be confirmed by work \cite{Whittle2009RELAX}:

\begin{enumerate}[(a)]
    \item
    The predicate appears after the modal verb, including words such as ``\emph{shall}'', ``\emph{must}'', ``\emph{can}''.
    \item
    The conditional adverbial clause usually begins with the word ``\emph{when}'' or ``\emph{if}''.
    \item
    The pronoun words, such as ``\emph{it}'' or ``\emph{them}'', usually do not appear in NL requirements.
\end{enumerate}

We assume that the NL requirements satisfy the above three conditions.
For the very few requirements violating them, we manually adjust the statements before running our algorithm to identify the semantic elements.

In the following sections, we introduce the heuristic rules and the related identification algorithm for each of the semantic elements, in line with the sequence in \textbf{Alg.} \ref{alg:mainAlg}.

\subsubsection{Identifying the \textbf{Operation}}
\label{subsubsec:opeIdentify}

\texttt{Operation} is the core action that \texttt{agent} should execute, and is usually the \emph{predicate} of the main clause of NL requirement.
Among all of the semantic elements, the \emph{operation} identification is the easiest since the \emph{Shall}-style statements have very obvious feature about the position of predicate verb.
Therefore our algorithm starts from the \emph{operation}.

During the requirement work-through, we find that there are different execution timing of \texttt{operation}.
Some requirements are with the phrases like ``\emph{have the ability to do something}''. That means, when the triggering \texttt{event} occurs, \texttt{operation} may not be executed, instead, it only has the possibility of execution.
And some requirements contain the phrases like ``\emph{not do something}'', that is, when the \emph{event} occurs, \texttt{operation} should not be executed.
Other requirements are with the meaning of ``\emph{should do something}''.
That means, when the \texttt{event} occurs, \texttt{operation} should be executed.
Therefore, we define three types of \texttt{operation\_mode}: \emph{ABLE}, \emph{NOT} and the \emph{default}.

The rules for identifying the \emph{operation} are as follows:
\begin{enumerate}[(a)]
    \item
    The \emph{predicate verb} in the main clause should become \texttt{operation}.
    \item The \texttt{operation} should belong to one of the above three categories and the rules for identifying these categories are as follows:
        \begin{enumerate}[(i)]
            \item
            If requirements' main clauses contain any directive, such as ``\emph{enable}'', ``\emph{be able to}'' and so on, \texttt{operation\_mode} should be \emph{ABLE}.
            For instance, the \texttt{operation} of the requirement ``The DronologyRuntimeMonitor shall be able to receive messages from any Dronology component'' is ``\emph{ABLE receive}''.
            \item
            If requirements' main clause contains a negative word, such as ``\emph{not}'', before the predicate verb, \texttt{operation\_mode} should be \emph{NOT}.
            For instance, the \texttt{operation} of ``When a UAV has an active onboard Obstacle Avoidance, the ObstacleAvoidance system shall not issue directives'' is ``\emph{NOT issue}''.
            \item
            Other requirements' \texttt{operation\_mode} is the \emph{default} form.
        \end{enumerate}
    \item
    The infinitive structure directly related to the \emph{predicate verb} of the main clause can be considered as the complement to it, since it usually expresses the purpose of \texttt{operation}.
    Therefore, the infinitive structure should be included in \texttt{operation}.
    For instance, for the requirement ``The RouteCreationUI shall allow a user to delete a route'', the \texttt{operation} is ``\emph{allow to delete}'.
    \item
    Some requirements are with copula verbs, usually ``\emph{be}'' and predicative.
    The composition of ``\emph{be}'' and predicative is regarded as \texttt{operation} with the purpose of expressing the relatively complete semantic.
    Also the predicative determines the condition of \texttt{agent}.
    For instance, the \texttt{operation} of ``Only one instance of each registered drone shall be active at any time'' is ``\emph{be active}''.
\end{enumerate}

The algorithm for \emph{operation} identification based on the above rules is depicted in \textbf{Alg.} \ref{alg: OperationIdentify}.

\begin{algorithm}[!htb]
    \SetKwComment{command}{right mark}{left mark}
    \caption{OperationIdentification}
    \label{alg: OperationIdentify}
    \KwIn {clause (single requirement entry), pos (result of POS), dependencies (result of SDP)}
    \KwOut {operation}
    \uIf {\rm ``\emph{shall}'' $\in$ clause \textbf{or} ``\emph{must}'' $\in$ clause} {
        operation = the word following ``\emph{shall}'' or ``\emph{must}''
    }
    \uElseIf {\rm ``\emph{can}'' $\in$ clause \textbf{or} ``\emph{may}'' $\in$ clause} {
        operation = the word following ``\emph{can}'' or ``\emph{may}''\\
        operation.operation\_mode = \emph{ABLE}
    }
    \tcc{If the first word of the clause is past participle or present participle of a verb, the verb is operation}
    \uElseIf {\rm pos\textbf{[}first\_word\textbf{]} == \emph{VBN} or \emph{VBG}} {
        operation = the first word of this clause
    }
    \tcc{If ``be'' appears, the clause is SVC format or passive voice}
    \ElseIf {\rm ``\emph{be}'' $\in$ clause} {
        \uIf {\rm $\exists$ dep $\in$ dependencies satisfies: (dep.type == \emph{cop} \textbf{and} dep.end == ``\emph{be}'')} {
            operation = link verb + copula
        }
        \ElseIf {\rm pos\textbf{(}word following ``be''\textbf{)} == \emph{VBN}} {
            operation = word following ``be''
        }
    }
    \tcc{If the clause has an infinitive structure directly related to predicate verb of the main clause}
    \If {\rm $\exists$ dep $\in$ dependencies satisfies: (dep.type == \emph{xcomp} \textbf{and} dep.start == operation \textbf{and} dep.end == ``\emph{to}'' \textbf{and} pos\textbf{(}word following ``to''\textbf{)} == \emph{VB})} {
        operation = operation + ``\emph{to}'' + word following ``to''
    }
    \tcc{If there is a negative word in the clause}
    \If {\rm $\exists$ dep $\in$ dependencies satisfies: (dep.type == \emph{neg} \textbf{and} dep.start == operation)} {
        operation.operation\_mode = \emph{NOT}
    }
    \textbf{return} operation
\end{algorithm}

\subsubsection{Identifying the \textbf{Agent}}

\texttt{Agent} is the executor of \texttt{operation}, and is usually the \emph{real subject} of the main clause.

The rules for identifying the \emph{agent} in NL requirements are as follows. 
\begin{enumerate}[(a)]
    \item
    For the requirements described in active voice, the real executor of the \emph{operation} is the \emph{sentence subject}.
    So the \emph{noun} that has \emph{nsubj} dependency with the \emph{predicate verb} of the main clause should become \texttt{agent}.
    Note: \emph{nsubj} from the SDP results of CoreNLP \cite{Marneffe2008Stanford}, means nominal subject.
    \item
    For the requirements described in passive voice,
    \begin{enumerate}[(i)]
        \item
        If there is a \emph{noun} after ``\emph{by}'', the \emph{noun} is regarded as the \texttt{agent}. Note that the noun here may be single word or phrase.
        \item
        If there is no \emph{noun} after ``\emph{by}'', we denote \texttt{agent} as $\varnothing$, meaning that any \texttt{entity} can be the \texttt{agent}.
    \end{enumerate}
\end{enumerate}

The algorithm upon these rules is listed in \textbf{Alg.} \ref{alg: agentParse}.

\begin{algorithm}[!htb]
    \SetKwComment{command}{right mark}{left mark}
    \caption{AgentIdentification}
    \label{alg: agentParse}
    \KwIn {operation, dependencies}
    \KwOut {agent}
    \uIf {\rm the requirement is passive voice} {
        \uIf {\rm $\exists$ dep $\in$ dependencies satisfies: (dep.type == \emph{dep} \textbf{and} dep.end == ``\emph{by}'')} {
            \tcc{The agent is the real subject of the sentence}
            agent = \emph{EntityParse}(dependency.start, dependencies)
        }
        \Else {
            \tcc{If real subject is not found, the agent is any entity}
            agent = $\varnothing$
        }
    }
    \tcc{The noun with \emph{nsubj} dependency of predicate verb with the main clause should become agent}
    \ElseIf {\rm $\exists$ dep $\in$ dependencies satisfies: (dep.type == \emph{nsubj} \textbf{and} dep.start == operation)} {
        agent = \emph{EntityParse}(dep.end, dependencies)\\
    }
    \textbf{return} agent
\end{algorithm}

\subsubsection{Identifying the \textbf{Event}}

\texttt{Event} is the precondition or trigger(s) of a requirement. It is usually the \emph{abverbial clause of condition} of the sentence.

For the sake of identifying the conflicts related with \texttt{event}, we divide the event clause into finer five-tuple: \{\texttt{agent}, \texttt{operation}, \texttt{input}, \texttt{output}, \texttt{restriction}\}.
The approach of identifying these elements are same with that of annotating the \texttt{agent}, \texttt{operation}, \texttt{input}, \texttt{output} and \texttt{restriction} in NL requirement.

We made three rules for the automated \texttt{event} identification.

\begin{enumerate}[(a)]
    \item
    The \emph{conditional adverbial clause} starting with specific connective words, such as ``\emph{when}'' or ``\emph{if}'', if existing, is the \texttt{event} of the requirement.
    And the clause begins with one connective word and ends with one specific punctuation, such as comma, period and semicolon.
    \item
    If there are at least two conditional adverbial clauses, and these clauses are connected with ``\emph{and}'' or ``\emph{or}'', the \texttt{event} should also includes multiple parts connected with ``\emph{and}'' or ``\emph{or}''. 
    \item
    If there is no conditional adverbial clause, \texttt{event} should be marked as $\varnothing$, meaning the \texttt{requirement} can be executed under any circumstances.
\end{enumerate}

The \texttt{event} identification algorithm is depicted in \textbf{Alg.} \ref{alg: eventParse}.

\begin{algorithm}[htbp]
    \SetKwComment{command}{right mark}{left mark}
    \caption{EventIdentification}
    \label{alg: eventParse}
    \KwIn {clause (single requirement entry), the phrases of \emph{agent}}
    \KwOut {the phrases of \emph{event}}
    \uIf {\rm there are ``\emph{when}'' or ``\emph{if}'' in clause} {
        event\_clause = the portion from ``\emph{when}'' or ``\emph{if}'' to \emph{agent} of clause\\
        (pos, dependencies) = \emph{CoreNLPResolve}(event\_clause)\\
        event.operation = \emph{OperationParse}(event\_clause, pos, dependencies)\\
        event.agent = \emph{AgentParse}(event.operation, dependencies)\\
        (event.input, event.output) = \emph{InputOutputParse}(operation, dependencies)\\
        event.restriction = \emph{RestrictionParse}(event\_clause, pos, dependencies)\\
        \textbf{return} (event.agent, event.operation, event.input, event.output, event.restriction)
    }
    \Else {
        \textbf{return} $\varnothing$
    }
\end{algorithm}

\subsubsection{Identifying the \textbf{Input \& Output}}

\texttt{Input} is all of the data that already exist before, and is required by \texttt{operation} for its execution.
\texttt{Output} is all of objects that can be created, destroyed or altered after executing the \texttt{operation}.
These two elements are identified at the same time because they are always intertwined with each other in the requirement statements.

We gave five rules for the automated \texttt{input \& output} identification.

\begin{enumerate}[(a)]
    \item
    The \emph{direct object} of the \emph{predicate verb} of the main clause should be included in \texttt{input} and \texttt{output}.
    In SDP, the relation between predicate and its direct object is \emph{dobj}.
    The direct object is usually the core object altered during the \texttt{operation} execution.
    For instance, for requirement ``When a flight plan is executed, the VehicleCore shall send the next waypoint to the UAV'', the direct object is ``\emph{the next waypoint}''.
    \item
    If a requirement statement is in passive voice, its \emph{formal subject} should be included in \texttt{input} and \texttt{output}.
    Typically, the formal subject is recognized as the subject of active voice requirement, that is, the \emph{noun} that has \emph{nsubj} dependency with the predicate verb.
    For instance, in requirement ``When the RealTimeFlightUI is loaded, a map shall be displayed'', the formal subject is ``\emph{a map}''.
    \item
    Nouns and noun phrases in the preposition-object compound and adverbial components should be included in \texttt{input}, since these nouns are also used during executing \texttt{operation}.
    And they are identified by \emph{nmod} dependency, excluding \emph{nmod:by}, \emph{nmod:poss}, \emph{nmod:of} and \emph{nmod:at}.
    For instance, \texttt{input} of the requirement ``The RealTimeFlightUI shall allow users to follow one or multiple UAVs on the map'' are ``\emph{one or multiple UAVs}''.
    \item
    If there is no \texttt{entity} in \texttt{input} or \texttt{output}, it should be denoted as $\varnothing$.
\end{enumerate}

The algorithm based on these rules for \texttt{input} \& \texttt{output} identification is given in \textbf{Alg.} \ref{alg:inputOutputIdentify}.

\begin{algorithm}[htbp]
    \SetKwComment{command}{right mark}{left mark}
    \caption{InputOutputIdentification}
    \label{alg:inputOutputIdentify}
    \KwIn {operation, dependencies}
    \KwOut {input, output}
    input = output = $\varnothing$\\
    \Else {
        \tcc{Formal subject in passive voice and direct object of predicate verb}
        \If {\rm clause is passive voice \textbf{and} $\exists$ dep $\in$ dependencies satisfies:(dep.type == \emph{nsubjpass} \textbf{and} dep.start == operation \textbf{or} dep.type == \emph{dobj} \textbf{and} dep.start == operation)} {
            entity = \emph{EntityParse}(dependency.end, dependencies)\\
            input = input $\cup$ \{entity\}\\
            output = output $\cup$ \{entity\}
        }
    }
    \For {\rm \textbf{all} dep $\in$ dependencies} {
        \tcc{Nouns and noun phrases in preposition-object components}
        \If {\rm dep.type == \emph{nmod} \textbf{and} dep.type $\notin$ {\emph{nmod:poss}, \emph{nmod:of}, \emph{nmod:agent}, \emph{nmod:by}} } {
            entity = \emph{EntityParse}(dep.end, dependencies)\\
            input = input $\cup$ \{entity\}
        }
    }
    \textbf{return} (input, output)
\end{algorithm}

\subsubsection{Identifying the \textbf{Restriction}}

\texttt{Restriction} is the constraints on the \texttt{operation}, including time, place, execution order, frequency, quantity of \texttt{operation}, and so on.

We made four rules to help the automated \texttt{restriction} identification.

\begin{enumerate}[(a)]
    \item
    The adverbs having \emph{advmod} dependency with the \emph{operation} verb should be included in \texttt{restriction}, which modify the execution of \texttt{operation}.
    It can be identified by the dependency \emph{advmod} between the adverbial word(s) and the predicate or a modifier word that it serves to modify.
    All of the \emph{execution time}, \emph{execution place} and \emph{execution order} of \texttt{operation} can be identified in this way.
    \item
    The common signs of \emph{frequency} of \texttt{operation} are \emph{every period n}, \emph{n times per period n} and \emph{n times a/an time unit}.
    For the first two forms, the \emph{frequency} can be identified from the pattern \emph{every + time unit} or ``per''.
    And for the last one, it can be identified from \emph{nmod:tmod} from the SDP results, which means \emph{temporal modifier}.
    \item
    The \emph{quantity} of \texttt{operation} means all of the number constraints that should be satisfied during operation execution.
    For instance, ``The SingleUAVFlightPlanScheduler shall only execute one flight plan at a time for each UAV'' has a quantity restriction ``\emph{only one at a time}''.
    And it can be identified by the dependency \emph{nmod:at} between \texttt{operation} and word ``\emph{time}''.
    \item
    If there is no item in \texttt{restriction}, it should be denoted as $\varnothing$.
\end{enumerate}

The algorithm for \texttt{restriction} identification is listed in \textbf{Alg.} \ref{alg:restrictionIden}.

\begin{algorithm}[htbp]
    \SetKwComment{command}{right mark}{left mark}
    \caption{RestrictionIdentification}
    \label{alg:restrictionIden}
    \KwIn {clause (single requirement entry), pos, dependencies}
    \KwOut {restriction}
    restriction = $\varnothing$\\
    \For {\rm \textbf{all} dep $\in$ dependencies} {
        \If {\rm dep.type == \emph{advmod} \textbf{and} dep.end $\notin$ \{``\emph{when}'', ``\emph{then}''\} } {
            \tcc{recognize numbers}
            \uIf {\rm dep.end == ``\emph{only}'' \textbf{and} $\exists$ word $\in$ clause satisfies: (pos\textbf{(}word\textbf{)} == \emph{CD})} {
                restriction = restriction $\cup$ (``\emph{only}'' + word)
            }
            \Else {
                restriction = restriction $\cup$ dep.end
            }
        }
        \tcc{Timely modifications should be included in restriction}
        \If {\rm dep.type == \emph{case} \textbf{and} dep.start == ``\emph{time}''} {
            restriction = restriction $\cup$ (the portion from dep.end to ``\emph{time}'' of clause)
        }
    }
    \textbf{return} restriction
\end{algorithm}

\subsection{Experimental evaluation on the semantic element identification}
\label{subsec:semanticAnnotationExp}

In this section we use three open requirement sets which are Unmanned Aerial Systems (UAS) \cite{Cleland2018Dronology}, OPENCOSS \footnote{\url{http://www.opencoss-project.eu}} and WorldVista \footnote{\url{http://coest.org/datasets}}, to evaluate the performance of our semantic element identifying algorithm.

The UAS set including 99 requirements, built by University of Notre Dame, describe the functions of UAV control system based on the template of EARS (Easy Approach to Requirements Syntax) \cite{Mavin2009Easy}.

OPENCOSS, i.e., an Open Platform for EvolutioNary Certification Of Safety-critical Systems for the railway, avionics and automotive markets, is an European large scale project dedicated to produce the first European-wide open safety certification platform. 
This set includes 110 requirements.

The WorldVista set including 117 requirements describes the functions of an electronic health record and health information system.

Just as the discussion in Section \ref{subsec:semanticElementIden}, we need to adjust the requirements firstly to ensure all of them meet the prerequisites about the NL requirements for our automated processing.
The adjustment including adding modal verbs, adding conditional adverbial clauses keywords (i.e., ``\emph{if}'' and ``\emph{when}'') and replacing pronouns to the corresponding nouns.
We would like to calculate the ratio of this kind of requirements requiring adjustments.
Thus we scanned all of the requirements in these three datasets, and calculated the number of requirements which are made the specific adjustment, shown in \textbf{Table} \ref{adjusted-number}.
According to the table, we can say the ratio of adjusted requirements is small (overall ratio is about 10.74\%) and this step is almost effortless.

\begin{table*}[htb]
    \centering
    \caption{The number of manual-adjusted requirements in each requirement set}
    \label{adjusted-number}
    \renewcommand\arraystretch{1.5}
    \begin{tabular}{cccccc}
    \hline
    Req. set            & \#Req.    & \#Adding modal verbs  & \#Adding conditional adverbial    & \#Replacing pronouns  & Overall \\
                        &           & (Proportion(\%))      & clauses keywords (Proportion(\%)) & (Proportion(\%))      & (Proportion(\%)) \\
    \hline
    \textbf{UAV}        & 99        & 4 (4.04)              & 3 (3.03)                          & 14 (14.14)            & 17 (17.17) \\
    \textbf{OpenCoss}   & 110       & 1 (0.91)              & 2 (1.82)                          & 2 (1.82)              & 5 (4.55) \\
    \textbf{WorldVista} & 117       & 1 (0.85)              & 8 (6.78)                          & 4 (3.39)              & 13 (11.02) \\
    \textbf{Overall}    & 326       & 6 (1.84)              & 13 (3.99)                         & 20 (6.13)             & 35 (10.74) \\
    \hline
    \end{tabular}
\end{table*}

Our testing set for the semantic element identification includes the all 99 requirements from Unmanned Aerial Systems (UAS) requirement set \cite{Cleland2018Dronology}, 30 randomly selected from OpenCoss requirements \cite{Arora2016Automated} and 50 randomly selected from WorldVista requirements.

With purposes of evaluation, we need to build the reference answer firstly.
Thus, we labeled the semantic elements for the all 179 requirements (i.e., 99 of UAS requirements, 30 of OpenCoss and 50 of WorldVista) manually.
Particularly, this manual annotation includes two rounds.
In the first round, the first two authors annotated the all 179 requirements independently.
They analyzed each requirement and labeled their semantic tuples.
Then in the second round, they checked and discussed their annotations face-to-face and made an agreement on the results one by one.
This manual annotation occurs before running our algorithm.
This process is not affected by the automated results, because we froze and mustn't modify the reference answer after our algorithm running.  

For measurement, we select the metrics of \emph{overall accuracy}, the \emph{accuracy} of each tuple and the \emph{average accuracy}.
The \emph{overall accuracy} refers to the proportion of requirements whose every semantic element is correctly identified.
And the \emph{accuracy} of each tuple refers to the proportion of requirements whose specific tuple is correctly identified, in all requirements.
And finally we calculate the \emph{average accuracy} on all tuples of the all requirements.
The result is shown in \textbf{Table} \ref{tab: semanticElemIden}.

\begin{table*}[htb]
    \centering
    \caption{Evaluation results of the automated semantic elements identification}
    \label{tab: semanticElemIden}
    \renewcommand\arraystretch{1.5}
    \begin{tabular}{cccccc}
    \hline
                                                &           & \textbf{UAV}  & \textbf{OpenCoss}  & \textbf{WorldVista}  & \textbf{Average} \\
    \hline
    \multicolumn{2}{c}{Total requirements(\#)}              & 99            & 30                 & 50                   & 179    \\
    \multicolumn{2}{c}{Correct annotated requirements (\#)} & 86            & 25                 & 31                   & 142    \\
    \multicolumn{2}{c}{Overall accuracy(\%)}                & 86.87         & 83.33              & 62.00                & 79.33  \\
    \hline
    \multirow{6}{*}{Accuracy of each tuple(\%)} & Event     & 98.99         & 96.97              & 98.00                & 97.97  \\
                                                & Agent     & 100.00        & 100.00             & 100.00               & 100.00 \\
                                                & Operation & 94.95         & 93.33              & 88.00                & 92.09  \\
                                                & Input     & 91.92         & 86.67              & 74.00                & 84.20  \\
                                                & Output    & 96.97         & 100.00             & 90.00                & 95.66  \\
                                              & Restriction & 98.99         & 100.00             & 100.00               & 99.66  \\
    \hline
    \multicolumn{2}{c}{Average accuracy of all tuples (\%)} & 96.97         & 96.16              & 91.67                & 94.93  \\
    \hline
    \end{tabular}
\end{table*}

From Table \ref{tab: semanticElemIden} we can make the following observations:

\begin{itemize}
    \item
    The \emph{overall accuracy} of the three datasets is between 62.0\% and 86.87\%.
    And the average of the \emph{overall accuracy} is 79.33\%. The identification on \emph{WorldVista} is the weakest.
    The primary reason is that comparing with the other two datasets, most of the requirements in \emph{WorldVista} represent more complicated sentence structure, like multiple nested clauses, lots of infinitives and enum items. 
    \item
    The \emph{average accuracy} of all tuples of the three datasets are between 91.67\% and 96.97\%.
    And the average of this metric on the three sets is 94.93\%.
    This means that our algorithm works great on identifying the semantic elements from NL requirements.
    \item
    Except \texttt{input}, the average accuracy of other tuples is above 92\%, and the accuracy of \texttt{agent} is 100\%.
    The excellent identification of \texttt{agent} is due to the well performance of CoreNLP \cite{CoreNLP2014} on identifying the subject of sentences.
    However, the accuracy of the \texttt{input} identification is lower than other tuples.
    With further analysis, we found that more than 80\% errors of \texttt{input} are caused by the false recognition of \texttt{modifier} in \texttt{entity} although the \texttt{base} is identified correctly.
    Since our algorithm for conflict detection mainly uses the \texttt{base} of \texttt{entity}, and \texttt{modifier}s only provide some assists (seen in Section \ref{sec:conflictDetection}),
    the errors of \texttt{input} don't cause much impact on the subsequent conflict detection, which can be seen in the experiment for evaluating the conflict detection algorithm (in Section \ref{subsec:conflictDetectionEva}).
\end{itemize}

\section{Phase II: Semantic annotation-based requirements conflict detection}
\label{sec:conflictDetection}

The purpose of this section is to detect the conflicts between the requirements which have been annotated with semantic elements.
To achieve this goal, we firstly define the types of requirements conflicts and give their detection rules.
Then, our conflict detection algorithm is purposed.

\subsection{Definition of the conflicts between software requirements}
\label{subsec:conflictDef}

The conflicts we concern can be referred as \emph{3Is}: \emph{Inconsistency}, \emph{Inclusion} and \emph{Interlock}.
These three categories can be further divided into seven finer types, which will be introduced in the following sections.

During describing the conflicts definition and their detection rules, three binary semantical relations between the tuples or their finer elements (i.e., string) are evolved: \emph{equivalent} (``=''), \emph{inclusion}(``$\supset$'') and \emph{contradiction}(``$\nparallel$'').
And the related atomic operators are given in the following Section \ref{subsec:atomicDef}.

\subsubsection{Requirement Inconsistency}
\label{subsubsec:reqInconsistencyDef}

\theoremstyle{definition}
\begin{definition}
	If two requirements $Req_1$ and $Req_2$ cannot be satisfied meanwhile, there is an \emph{inconsistency} relation between them.
\end{definition}

This kind of conflict gets the most common attention in the existing works.
Chentouf \cite{Chentouf2014Managing} proposed four incompatible types between requirements focusing on \texttt{operation} and \texttt{event}.
Kim et al. \cite{Kim2007Managing} proposed \emph{activity conflicts} and \emph{resource conflicts} based on the structuring Action (verb) + Object + Resource.
Moser et al. \cite{Moser2011Requirements} aimed to identify three conflict types, including conflicts between functional requirements, conflicts between requirement and requirement constraint, and conflicts between requirement and EBNF grammar.
However, Moser's work didn't give the specific types of conflicts between requirements.

Based on these work, we propose three conflicting types: \emph{operation-inconsistency}, \emph{restriction-inconsistency} and \emph{event-inconsistency}, which are mutually exclusive.

\begin{enumerate}[(1)]
    \item \emph{Operation Inconsistency}\\
    If and only if $Req_1$ and $Req_2$ meet:
    \begin{enumerate}[(a)]
        \item
        \texttt{event} and \texttt{agent} of $Req_1$ and those of $Req_2$ are equivalent;
        \item
        \texttt{input} of $Req_1$ contains that of $Req_2$, or \texttt{input} of $Req_2$ contains that of $Req_1$;
        \item
        \texttt{output} of $Req_1$ contains that of $Req_2$, or \texttt{output} of $Req_2$ contains that of $Req_1$;
        \item
        The \texttt{operation} of $Req_1$ contradicts with that of $Req_2$.
    \end{enumerate}
    It is said that there is an \emph{operation-inconsistency} relation between $Req_1$ and $Req_2$, which is an antisymmetric relation.
    The formal description is as follows:\\
    $Req_1.event$ $=$ $Req_2.event$ $\wedge$ $Req_1.agent$ $=$ $Req_2.agent$ $\wedge$ $Req_1.operation$ $\nparallel$ $Req_2.operation$ $\wedge$ $\textbf{(}$$(Req_1.input$ $\supset$ $Req_2.input$ $\wedge$ $Req_1.output$ $\supset$ $Req_2.output)$ $\vee$ $(Req_2.input$ $\supset$ $Req_1.input$ $\wedge$ $Req_2.output$ $\supset$ $Req_1.output)$$\textbf{)}$
    \\$\Leftrightarrow operation$-$inconsistency\ (Req_1, Req_2)$

    \item \emph{Restriction Inconsistency}\\
    If and only if $Req_1$ and $Req_2$ meet:
    \begin{enumerate}[(a)]
        \item
        All of \texttt{event}, \texttt{agent}, and \texttt{operation} of $Req_1$ are equal with those of $Req_2$ respectively;
        \item
        \texttt{input} of $Req_1$ contains that of $Req_2$, or \texttt{input} of $Req_2$ contains that of $Req_1$;
        \item
        \texttt{output} of $Req_1$ contains that of $Req_2$, or \texttt{output} of $Req_2$ contains that of $Req_1$;
        \item
        \texttt{restriction} of $Req_1$ contradicts with that of $Req_2$.
    \end{enumerate}
    It is said that there is a \emph{restriction-inconsistency} relation between $Req_1$ and $Req_2$, which is an symmetric relation.
    The formal description is as follows:\\
    $Req_1.event$ $=$ $Req_2.event$ $\wedge$ $Req_1.agent$ $=$ $Req_2.agent$ $\wedge$ $Req_1.operation$ $=$ $Req_2.operation$ $\wedge$ $Req_1.restriction$ $\nparallel$ $Req_2.restriction$ $\wedge$ $\textbf{(}$$(Req_1.input$ $\supset$ $Req_2.input$ $\wedge$ $Req_1.output$ $\supset$ $Req_2.output)$ $\vee$ $(Req_2.input$ $\supset$ $Req_1.input$ $\wedge$ $Req_2.output$ $\supset$ $Req_1.output)$$\textbf{)}$
    \\$\Leftrightarrow restriction$-$inconsistency\ (Req_1, Req_2)$

    \item \emph{Event Inconsistency}\\
    If and only if $Req_1$ and $Req_2$ meet:
    Once the \texttt{operation} of $Req_2$ is executed, the \texttt{event} of $Req_1$ cannot be satisfied which means that $Req_1$ would not be triggered. This can be described as: suppose there is a \texttt{condition} $\gamma$ of the \texttt{event} of $Req_1$ which satisfies:
    \begin{enumerate}[(a)]
        \item
        The \texttt{agent} of $\gamma$ is equal with the agent of $Req_2$, and the \texttt{operation} of $\gamma$ contradicts with that $Req_2$;
        \item
        The \texttt{input} and \texttt{output} of $\gamma$ contain (or are contained by) the corresponding tuple of $Req_2$.
    \end{enumerate}

    It is said that there is an \emph{event-inconsistency} relation between $Req_1$ and $Req_2$, and the formal description is as follows:\\
    $\exists$ $Condition$ $\in$ $Req_1.event,$ $Condition.agent$ $=$ $Req_2.agent$ $\wedge$ $Condition.operation$ $\nparallel$ $Req_2. operation$ $\wedge$ $\textbf{(}$$(Condition.input$ $\supset$ $Req_2.input$ $\wedge$ $Condition.output$ $\supset$ $Req_2.output)$ $\vee$
    $(Req_2.input$ $\supset$ $Condition.input$ $\wedge$ $Req_2.output$ $\supset$ $Condition.output)$$\textbf{)}$
    \\$\Leftrightarrow event$-$inconsistency\ (Req_1, Req_2)$
\end{enumerate}

\subsubsection{Requirement Inclusion}
\label{subsubsec:reqInclusionDef}

\theoremstyle{definition}
\begin{definition}
	If some tuples of $Req_1$ semantically contain the corresponding tuples of $Req_2$, and the other tuples are semantically equal, then there is an \emph{inclusion} relation between $Req_1$ and $Req_2$.
\end{definition}

According to the above definition, $Req_2$ is essentially a (partial) redundant requirement.
It should be removed (sometimes $Req_1$ needs modification meanwhile) to optimize the requirement set, since inconsistence will occur if $Req_2$ is changed while $Req_1$ not.
Therefore, we consider \emph{inclusion} as an conflict category.

In the existing research, Walter et al. \cite{Walter2017formalization} defined the type of \emph{requirement redundancy} and Chentouf \cite{Chentouf2014Managing} focused on one special type of inclusion: duplicated requirement.
This duplicated requirements are relatively easy to detect. Besides duplication, we concern the semantically containing relation too. We defined two finer \emph{inclusion} types: \emph{operation-inclusion} and \emph{event-inclusion}, which are mutually exclusive.

\begin{enumerate}[(1)]
    \item \emph{Operation Inclusion}\\
    If and only if $Req_1$ and $Req_2$ meet:
    \begin{enumerate}[(a)]
        \item
        \texttt{event} and \texttt{agent} of $Req_1$ are equal with those of $Req_2$ respectively;
        \item
        All of the \texttt{operation}, \texttt{input}, \texttt{output} and \texttt{restriction} of $Req_1$ contain the corresponding tuples of $Req_2$.
    \end{enumerate}

    It is said that there is an \emph{operation-inclusion} relation between $Req_1$ and $Req_2$, which is an antisymmetric relation.
    The formal description is as follows:\\
    $Req_1.event$ $=$ $Req_2.event$ $\wedge$ $Req_1.agent$ $=$ $Req_2.agent$ $\wedge$ $Req_1.operation$ $\supset$ $Req_2.operation$
    $\wedge$ $Req_1.input$ $\supset$ $Req_2.input$ $\wedge$ $Req_1.output$ $\supset$ $Req_2.output$ $\wedge$ $Req_1.restriction$ $\supset$ $Req_2.restriction$
    \\$\Leftrightarrow operation$-$inclusion\ (Req_1, Req_2)$

    \item \emph{Event Inclusion}\\
    If and only if $Req_1$ and $Req_2$ meet:
    \begin{enumerate}[(a)]
        \item
        As long as \texttt{event} of $Req_1$ is satisfied, \texttt{event} of $Req_2$ can be satisfied for sure, which means that \texttt{event} of $Req_1$ contains that of $Req_2$.
        \item
        Each of \texttt{agent}, \texttt{operation}, \texttt{input}, \texttt{output}, and \texttt{restriction} of $Req_1$ is equal with the corresponding tuple of $Req_2$ respectively.
    \end{enumerate}

    It is said that there is an \emph{event-inclusion} relation between $Req_1$ and $Req_2$, which is an antisymmetric relation.
    The formal description is as follows:\\
    $Req_1.event$ $\supset$ $Req_2.event$ $\wedge$ $Req_1.agent$ $=$ $Req_2.agent$ $\wedge$ $Req_1.operation$ $=$ $Req_2.operation$
    $\wedge$ $Req_1.input$ $=$ $Req_2.input$ $\wedge$ $Req_1.output$ $=$ $Req_2.output$ $\wedge$ $Req_1.restriction$ $=$ $Req_2.restriction$
    \\$\Leftrightarrow event$-$inclusion\ (Req_1, Req_2)$
\end{enumerate}

\subsubsection{Requirement Interlock}
\label{subsubsec:reqInterlockDef}

The interlock relation between requirements are described upon the \emph{requirement interlock graph}.

\theoremstyle{definition}
\begin{definition}
    The \emph{requirement interlock graph} is a digraph, in which each requirement is a vertex. If two requirements meet certain \emph{dependency} relation, there is an edge between them.
\end{definition}

\begin{definition}
	 If there is a circuit in the interlock graph, we say there is an \emph{interlock} relation among the requirements in the circuit.
\end{definition}

The circuit indicates the interdependency between several requirements, which will probably cause the instability and unreliability during requirement-oriented development.
Therefore, we consider \emph{interlock} is a category of conflicts.

We define two types of interlock: \emph{operation-event-interlock} and \emph{input-output-interlock}. 

\begin{enumerate}[(1)]
    \item \emph{Operation-Event Interlock}\\
    $Req_1$ and $Req_2$ has \emph{operation-event-dependency} relation if as long as the \texttt{operation} of $Req_1$ is executed, the \texttt{event} of $Req_2$ must be triggered.
    And there will be an edge from $Req_1$ to $Req_2$ in the interlock graph.
    
    The formal description of the condition of \emph{operation-event-dependency} is as follows:\\
    $\forall$ $Event$ $\in$ $Req_2.event,$ $(Req_1.agent$ $=$ $Event.agent$ $\wedge$ $Req_1.operation$ $\supset$ $Event.operation$
    $\wedge$ $Event.restriction$ $=$ $Req_1.restriction$ $\wedge$ $Req_1.input$ $\supset$ $Event.input$ $\wedge$ $Req_1.output$ $\supset$ $Event.output)$
    \\$\Leftrightarrow operation$-$event$-$dependency\ (Req_1, Req_2)$

	If there are multiple requirements with the \emph{operation-event-dependency} relation in a circle of the requirement interlock graph, we say that all these requirements have \emph{operation-event-interlock} relation.

    Due to the \texttt{event} of these requirements can be triggered in any circumstances, they will be executed repeatedly and forever.
    There are likely potential conflicts among these requirements.

    \item \emph{Input-Output Interlock}\\
    $Req_1$ and $Req_2$ has \emph{input-output-dependency} relation if the \texttt{output} of $Req_1$ contains the \texttt{input} of $Req_2$ and there are no \emph{event inconsistency} between these two requirements.
    And there is an edge from $Req_1$ to $Req_2$ in the interlock graph.
    
    The formal description of the condition of \emph{input-output-dependency} is as follows:\\
    $(\exists$ $Entity_1$ $\in$ $Req_1.output,$ $\exists$ $Entity_2$ $\in$ $Req_2.input,$ $Entity_1$ $\supset$ $Entity_2)$ $\wedge$ $\neg$ $event$-$inconsistency\ (Req_1, Req_2)$ $\wedge$ $\neg$ $event$-$inconsistency\ (Req_2, Req_1)$
    \\$\Leftrightarrow input$-$output$-$dependency\ (Req_1, Req_2)$ 
    
    If there is a circuit in the graph and all requirements in this circle have \emph{input-output-dependency}, these requirements are said to have \emph{input-output-interlock} relation.
    
    During the execution, \texttt{input} of each requirement will change continuously, which may result in the unstable implementation.
\end{enumerate}

\subsection{Definition of atomic operators}
\label{subsec:atomicDef}

Three binary atomic operators (i.e., \emph{equivalent} (``=''), \emph{inclusion}(``$\supset$'') and \emph{contradiction}(``$\nparallel$'')) are used in the definition of conflicts and their definitions are critical for understanding and detecting the conflicts, undoubtedly.

There are six elements with these three operators, including \texttt{string}, \texttt{entity}, \texttt{entity\_set}, \texttt{operation}, \texttt{restriction} and \texttt{event}.
We introduce these six elements and their detection rules in the three operators in the following sections.

\subsubsection{String}
According to the BNF definition of requirement in Section \ref{subsec:semanticDefinition}, the most basic representation of the tuples including \texttt{predicate}, \texttt{modifier} and \texttt{base} are in the form of \texttt{strings}.
We only consider \emph{equivalent} relation between \texttt{strings}, which means that two \texttt{strings} are identical in case insensitive or synonyms.

\subsubsection{Entity}
According to the BNF in Section \ref{subsec:semanticDefinition}, the entity is primarily composed of \texttt{base} and \texttt{modifier}.
We take $Entity_1$ and $Entity_2$ as an example to describe the definition of the three atomic operators upon \texttt{entity}.

The specific conditions are given below:

\begin{enumerate}[(a)]
    \item Inclusion:
    \begin{enumerate}[(i)]
        \item
        When $Entity_1$ denotes any kind of entity and $Entity_2$ denotes some special kind of entity, then there is an \emph{inclusion} relation from $Entity_1$ to $Entity_2$, such as ``\emph{UAV}'' $\supset$ ``\emph{UAV in flight}''.
        The formal checking condition can be described as:\\
        $Entity_1.base$ $=$ $Entity_2.base$ $\wedge$ $(\forall$ $string_1$ $\in$ $Entity_1.modifier,$ $\exists$ $string_2$ $\in$ $Entity_2.modifier,$ $string_1$ $=$ $string_2)$
        \\$\Rightarrow\ Entity_1 \supset Entity_2$

        \item
        In some cases, $Entity_2$ denotes an entire entity, while $Entity_1$ denotes a part of it or a parameter of it, then there is an \emph{inclusion} relation from $Entity_1$ to $Entity_2$, such as ``\emph{UAV}'' $\supset$ ``\emph{wings of UAV}''.
        In this study two representations of the part relation are involved: ``\emph{of}'' and  possessive noun (i.e., adding apostrophe ('s ) to a noun).\\
        $(\exists$ $string$ $\in$ $Entity_2.modifier,$ $string$ $=$ $``of$ $Entity_1.base"$ $\vee$ $string$ $=$ $``Entity_1.base's")$
        \\$\Rightarrow\ Entity_1 \supset Entity_2$
    \end{enumerate}

    \item Equivalent:
    \begin{enumerate}[(i)]
        \item If both the \texttt{base} and \texttt{modifier} of $Entity_1$ and $Entity_2$ are equal, there is an \emph{equivalent} relation between them.\\
        $Entity_1.base$ $=$ $Entity_2.base$ $\wedge$ $Entity_1.modifier$ $\in$ $Entity_2.modifier$ $\wedge$ $Entity_2.modifier$ $\in$ $Entity_1.modifier$
        \\$\Rightarrow\ Entity_1 = Entity_2$
    \end{enumerate}
\end{enumerate}

\subsubsection{Entity\_set}
\texttt{Entity\_set} is a set of \texttt{entity}.

In some cases, $Entity\_set_1$ semantically includes all entities of $Entity\_set_2$.
Then, there is an unidirectional \emph{inclusion} relation from $Entity\_set_1$ to $Entity\_set_2$.

In some more particular cases, all of \texttt{entity} of $Entity\_set_1$ are semantically equal to those of $Entity\_set_2$.
Then, there is an \emph{equivalent} relation between these two \texttt{entity\_set}.

The specific conditions are given below:

\begin{enumerate}[(a)]
    \item Inclusion:
    each \texttt{entity} in $Entity\_set_2$ is included in an \texttt{entity} in $Entity\_set_1$.\\
    $\forall$ $entity_2$ $\in$ $Entity\_set_2,$ $\exists$ $entity_1$ $\in$ $Entity\_set_1,$ $entity_1$ $\supset$ $entity_2$
    \\$\Rightarrow\ Entity\_set_1 \supset Entity\_set_2$

    \item Equivalent:
    $Entity\_set_1$ and $Entity\_set_2$ have mutual \emph{inclusion} relation.\\
    $Entity\_set_1$ $\supset$ $Entity\_set_2$ $\wedge$ $Entity\_set_2$ $\supset$ $Entity\_set_1$
    \\$\Rightarrow\ Entity\_set_1 = Entity\_set_2$
\end{enumerate}

\subsubsection{Operation}
We take $Operation_1$ and $Operation_2$ as the example.

In some cases, $Operation_1$ semantically contains $Operation_2$, that is, the execution of $Operation_1$ also means the execution of $Operation_2$.
Then there is an unidirectional \emph{inclusion} relation from $Operation_1$ to $Operation_2$.

In some more particular cases, $Operation_1$ and $Operation_2$ are exactly the same.
Then there is a symmetric \emph{equivalent} relation between them.

In some other cases, $Operation_1$ and $Operation_2$ are semantically contradictory, that is, they cannot be executed at the same time.
Then there is a symmetric \emph{contradiction} relation between them.

The specific conditions are given below:

\begin{enumerate}[(a)]
    \item Equivalent:\\
    $Operation_1.operation\_mode$$=$$Operation_2.operation\_mode$ $\wedge$ $Operation_1.predicate$$=$$Operation_2.predicate$
    \\$\Rightarrow\ Operation_1 = Operation_2$
    \\$\Rightarrow\ Operation_1 \supset Operation_2$ 
    
    \item Inclusion:
    $Operation_1$ includes $Operation_2$ if  the \texttt{predicate} of $Operation_1$ is equal with that of $Operation_2$, and the $opeartion\_mode$ of $Operation_1$ is \emph{default}(i.e., do) or \emph{NOT}, and the $opeartion\_mode$ of $Operation_2$ is \emph{ABLE}.\\
    We make this definition because the mode of \emph{ABLE} means the operation may or may not be executed, while \emph{default} or \emph{NOT} means operation should or not be executed, contains the meaning of \emph{ABLE}.\\
    $Operation_1.predicate$ $=$ $Operation_2.predicate$ $\wedge$ $Operation_1.operation\_mode$ $=$ ``Not'' $or$ $\varnothing$ $\wedge$ $Operation_2.operation\_mode$ $=$ ``Able''
    \\$\Rightarrow\ Operation_1 \supset Operation_2$
    
    \item Contradiction:
    $Operation_1$ contradicts with $Operation_2$ if the \texttt{predicate} of $Operation_1$ is equal with that of $Operation_2$, and one $opeartion\_mode$ is \emph{default} (i.e., do) and the other is \emph{NOT}.\\
    $Operation_1.predicate$ $=$ $Operation_2.predicate$ $\wedge$ $\textbf{(}$$(Operation_1.operation\_mode$ $=$ ``Not'' $\wedge$ $Operation_2.operation\_mode$ $=$ $\varnothing)$ $\vee$ $(Operation_1.operation\_mode$ $=$ $\varnothing$ $\wedge$ $Operation_2.operation\_mode$ $=$ ``Not''$)\textbf{)}$ 
    \\$\Rightarrow\ Operation_1 \nparallel Operation_2$
\end{enumerate}

\subsubsection{Restriction}
We take $Restriction_1$ and $Restriction_2$ as an example.

$Restriction_1$ contains $Restriction_2$, which means that $Restriction_1$ is stricter than $Restriction_2$.
And when $Restriction_1$ is satisfied, $Restriction_2$ also can be satisfied.
Then there is an unidirectional \emph{inclusion} relation from $Restriction_1$ to $Restriction_2$.

In some more particular cases, $Restriction_1$ and $Restriction_2$ are exactly the same.
Then there is a symmetric \emph{equivalent} relation between them.

The specific conditions are given below:

\begin{enumerate}[(a)]
    \item Inclusion:
    all of constraints in $Restriction_2$ are also in $Restriction_1$.\\
    $\forall$ $string_2$ $\in$ $Restriction_2,$ $\exists$ $string_1$ $\in$ $Restriction_1,$ $string_1$ $=$ $string_2$
    \\$\Rightarrow\ Restriction_1 \supset Restriction_2$

    \item Equivalent:
    $Restriction_1$ and $Restriction_2$ are exactly the same.\\
    $Restriction_1$ $\supset$ $Restriction_2$ $\wedge$ $Restriction_2$ $\supset$ $Restriction_1$
    \\$\Rightarrow\ Restriction_1 = Restriction_2$
\end{enumerate}

\subsubsection{Event}
We take $Event_1$ and $Event_2$ as an example.

$Event_1$ containing $Event_2$ means when $Event_1$ is satisfied, $Event_2$ must also be satisfied.
Then there is an unidirectional \emph{inclusion} relation from $Event_1$ to $Event_2$.

When $Event_1$ and $Event_2$ are exactly the same, or the corresponding tuples are synonymous, there is a symmetric \emph{equivalent} relation between them.

If any two \texttt{conditions} of one \texttt{event} are semantically contradictory, then this \texttt{event} cannot be satisfied.
We say that there is a symmetric \emph{contradiction} relation between these two \texttt{conditions}.

The specific conditions are given below:

\begin{enumerate}[(a)]
    \item Inclusion:
    if for any condition $\gamma$ in $event_2$, there are at least one conditions in $event_1$ that can contain $\gamma$, $event_1$ contains $event_2$.\\
    $\forall$ $condition_2$ $\in$ $event_2,$ $\exists$ $condition_1$ $\in$ $event_1,$ $condition_1.agent$ $=$ $condition_2.agent$ $\wedge$ $condition_1.operation$ $\supset$ $condition_2.operation$ $\wedge$ $condition_1.input$ $\supset$ $condition_2.input$ $\wedge$ $condition_1.output$ $\supset$ $condition_2.output$ $\wedge$ $condition_1.restriction$ $=$ $condition_2.restriction$
    \\$\Leftrightarrow\ event_1 \supset event_2$
    
    \item Contradiction:
    if any two conditions of one \texttt{event} conflict with each other, this \texttt{event} is contradictory.\\
    $\exists$ $condition_1,$ $condition_2$ $\in$ $event,$ $condition_1.agent$ $=$ $condition_2.agent$ $\wedge$ $condition_1.input$ $\supset$ $condition_2.input$ $\wedge$ $condition_1.operation$ $\nparallel$ $condition_2.operation$ $\wedge$ $condition_1.output$ $\supset$ $condition_2.output$
    \\$\Leftrightarrow\ event \nparallel event$
    
    \item Equivalent:
    $Event_1$ and $Event_2$ are exactly same or the corresponding tuples are synonymous, there are \emph{equivalent} relation between them.\\
    $event_1$ $\supset$ $event_2$ $\wedge$ $event_1$ $\supset$ $event_2$
    \\$\Leftrightarrow\ event_1 = event_2$
\end{enumerate}

\subsection{Software requirements conflict detecting algorithm}
\label{subsec: conflictDetectionAlg}

Before conducting the algorithm of conflict detection based on their definitions, we design four steps for the preprocessing.

\begin{enumerate}[(1)]
    \item
    Object clause is often a grammar-complete sentence. For the sake of conflict identification, we parse object clause into eight-tuple too. 
    The object clause can be automatically identified by \emph{ccomp} dependency.
    NOTE: \emph{ccomp} from the SDP, means clausal complement. 
    \item
    Detect contradiction relation of the multiple \texttt{event}s of one single requirement, namely self-contradictory \texttt{event}.
    \item
    Split the requirement statements.
    If one requirement has multiple event clauses with ``or'' relation, combine each clause and the main clause into new requirements. Remove the original requirements from the requirement set.
    \item
    Put the requirements that have the same \texttt{event} and \texttt{agent} into a group.
    Then we can identify the conflicts between requirements according to their groups.
    Some conflicts can only occur within a group such as \emph{operation-include}, \emph{operation-inconsistency} and \emph{restriction-inconsistency}.
    And \emph{event-include} and \emph{event-inconsistency} happen in the requirements in different groups.
    Requirements grouping is dedicated to improve the efficiency of the conflict detection.
\end{enumerate}

Then we detect the conflicts according to the definitions in Section \ref{subsec:conflictDef}. The detection algorithm is listed in \textbf{Alg.} \ref{alg:conflictDetect}

\begin{algorithm}[htbp]
    \SetKwComment{command}{right mark}{left mark}
    \caption{Conflict Detection Algorithm}
    \label{alg:conflictDetect}
    \KwIn {A requirement set $R$ with semantically annoated requirements $r_1$, $r_2$, $\dots$, $r_n$}
    \KwOut {The conflict set $C$}
    \tcc{Preparing for conflicts detection}
    \For {\rm \textbf{all} requirement $r$ $\in$ $R$} {
        \If {\rm there is an object clause in $r$} {
            Replace $r$ with object clause in $R$.
        }
      	\If {\rm \texttt{event} $e$ of $r$ is self-contradictory, namely \textbf{conflict}($e$, $e$)} {
      		Add ($r$, $r$) to $C$.\\
      	    Remove $r$ from $R$.
        }
        \If {\rm \texttt{event} $e$ of $r$ $\neq$ \textbf{ALL} and \rm $e$ is consisted of multiple parts with ``or'' relation} {
            Combine each part of $e$ and the remaining part of $r$ into new requirements $r_1'$, $r_2'$, $\dots$.\\
            Add $r_1'$, $r_2'$, $\dots$ to $R$.\\
            Remove $r$ from $R$.
        }
    }
    \tcc{Detect the conflicts between the requirements in the same group}
    \For {\rm \textbf{all} pairs of requirements in the same group ($r_1$, $r_2$) $\in$ $R^2$} {
        \If {\rm \emph{operation-include}($r_1$, $r_2$) \textbf{or} \emph{operation-inconsistency}($r_1$, $r_2$) \textbf{or} \emph{restriction-inconsistency}($r_1$, $r_2$)} {
            Add ($r_1$, $r_2$) to $C$.
        }
    }
    \tcc{Detect other conflicts, and build interlock graphs}
    \For {\rm \textbf{all} pairs of requirements not in the same group ($r_1$, $r_2$) $\in$ $R^2$} {
        \If {\rm \emph{event-include}($r_1$, $r_2$) \textbf{or} \emph{event-inconsistency}($r_1$, $r_2$)} {
            Add ($r_1$, $r_2$) to $C$.
        }
        \If {\rm \emph{operation-event-dependency}($r_1$, $r_2$)} {
            Add edge($r_1$, $r_2$) to \emph{operation-event-interlock-graph}.
        }
    }
    \For {\rm \textbf{all} pairs requirements ($r_1$, $r_2$) $\in$ $R^2$} {
        \If {\rm \emph{input-output-dependency}($r_1$, $r_2$)} {
            Add edge($r_1$, $r_2$) to \emph{input-output-interlock-graph}.
        }
    }
    \tcc{Detect requirement interlocks}
    \For {\rm \textbf{all} circuit in \emph{operation-event-interlock-graph} or \emph{input-output-interlock-graph}} {
        Add all requirements in circuit ($r_1$, $r_2$, $\dots$, $r_n$) to $C$.
    }
    \textbf{return} $C$.
\end{algorithm}

Now we analyze the time complexity of the conflict detection algorithm.
Let the number of requirements in the input requirement set $R$ is $n_0$.
We traverse all the requirements in the first 14 lines, obviously, the time complexity is $O(n_0)$.

And after this, let the number of requirements in $R$ is $n$, therefore, the number of all pairs of requirements is $n^2$.
Then we traverse all the pairs by the same group (line 15 - 19), by not the same group (line 20 - 27), and the whole (line 28- 32).
Therefore, the time complexity is $O(n^2)$.

In the end, we find all the circuits from the graph with $n$ vertexes.
Let the number of edges is $e$, the time complexity is $O(ne)$.

In summary, the overall time complexity of the algorithm is $O(n_0 + n^2 + ne)$.
In the case where $n$ is closer to $n_0$, the time complexity is $O(n^2 + ne)$.
Considering that there are few conflicts in the requirements, therefore, $e$ is not much bigger than $n$.
The overall complexity is the square of the number of requirement.

\subsection{Experimental evaluation on the conflict detection}
\label{subsec:conflictDetectionEva}

To evaluate the performance of automated conflict detection, we used UAV and WorldVista requirement sets in Section \ref{subsec:semanticAnnotationExp}.
Different from that in the experiment of Section \ref{subsec:semanticAnnotationExp}, we used the full sets, that is, 99 requirements in UAV set and 117 in WorldVista set.
However, the conflicts among them are unknown in advance.

Therefore, we also selected the requirement sets of two more real projects whose conflicts can be collected meanwhile.
The first set is the requirements of a \emph{telecom management system} from work \cite{Chentouf2014Managing} and all of the conflicting requirements as well as the conflict types are given.

The second set includes the requirements of a \emph{Solar power supply system}\footnote{\url{http://www.caiso.com/pages/recent-documents.aspx/}} built by California ISO company since there are several versions of the requirement specifications online and the modified content are clearly annotated in the newer version.
These modifications are good source of conflicts among requirements cross versions.
For instance, one requirement is ``Market system shall calculate the total EIM Transfer Limits for both the import and export directions'' in the old version, and is modified into ``Market system shall calculate and broadcast the total EIM Transfer Limits for both the import and export directions'' in the new version.
Therefore, there is \emph{Operation Inclusion} between these two requirements cross two versions.
We selected 26 NL requirements from the latest three versions.
The other ones are with restricted NL (e.g., use case), not the target of this study.
We found 12 conflicting requirements with 7 conflicts, 5 \emph{Input-Output Interlock}, one \emph{Restriction Inconsistency} and one \emph{Operation Inclusion}.

For the \emph{UAV} and \emph{WorldVista} sets, we manually checked each of the detected results, judged and recorded their correctness or not.
Accordingly, we calculated the \emph{precision}. \emph{Recall} was not computed because we don't have the full sets of conflicts.
While, for \emph{Telecom} and \emph{Solar} we work out both the \emph{precision} and \emph{recall} by comparing the automated detection with the golden standard.
The result is shown in \textbf{Table} \ref{conflict_result}.
In this table, we list the total number of requirements in the four datasets, the number of conflicts (if known), the number of detected conflicts, number of the correct detection, as well as the precision and recall values.
Besides, we give the overall calculation of each metric.

\begin{table*}[htb]
	\centering
	\renewcommand\arraystretch{1.5}
    \caption{The evaluation results of our conflict detection algorithm on the four requirement sets}
    \label{conflict_result}
    \begin{tabular}{cccc|ccc}
    \hline
                            & \textbf{UAV} & \textbf{WorldVista} & \textbf{Overall} & \textbf{Telecom Management} & \textbf{Solar Power Supply} & \textbf{Overall} \\
    \hline
    Total requirements (\#) & 99           & 117                 & 216              & 14                          & 12                          & 26 \\
    Known conflicts (\#)    & ——           & ——                  & ——               & 7                           & 7                           & 14 \\
    Detected conflicts (\#) & 14           & 19                  & 33               & 8                           & 7                           & 15 \\
    Correct detected (\#)   & 11           & 15                  & 26               & 7                           & 7                           & 14 \\
    Precision (\%)          & 78.57        & 78.95               & 78.79            & 87.50                       & 100.00                      & 93.33 \\
    Recall (\%)             & ——           & ——                  & ——               & 100.00                      & 100.00                      & 100.00 \\
    \hline
    \end{tabular}
\end{table*}

\begin{table*}[htb]
    \centering
    \renewcommand\arraystretch{1.5}
    \caption{The performance of our approach on detecting the related conflict types in UAV and WorldVista}
    \label{tab:analysisOnConflictTypeDetection}
    \begin{tabular}{cccc}
    \hline
                        & \textbf{Operation inclusion} & \textbf{Restriction inconsistency} & \textbf{Entity interlock} \\
                        & CorrectDetect/Detected       & CorrectDetect/Detected             & CorrectDetect/Detected \\
    \hline
    \textbf{UAV}        & 0/0                          & 1/1                                & 10/13 \\
    \textbf{WorldVista} & 2/2                          & 0/0                                & 13/17 \\
    \hline
    \end{tabular}
\end{table*}

Overall, our algorithm yields great performance of precision and recall on the two datasets.
For the requirement sets with unknown conflicts (i.e., \emph{UAV} and \emph{WorldVista}), the precision is between 78\% and 79\%.
And for the other two sets with known conflicts (i.e., \emph{Telecom Management} and \emph{Solar Power Supply}), the precision is above 87\%, especially for \emph{Solar} the precision reaches 100\%.
And the detection recall on both these two datasets reaches 100\%.

For the sake of clearly evaluating the performance of our algorithm on the conflict type detection, we further analyzed the number of each type of conflicts detected by our algorithm in the sets of \emph{UAV} and \emph{WorldVista}, and the results are shown in \textbf{Table} \ref{tab:analysisOnConflictTypeDetection}.
We found that for the two requirement sets with unknown conflicts, the detected conflicts are mainly \emph{input-output-interlock}.
There are 13 and 17 \emph{input-output-interlock} conflicts in \emph{UAV} and \emph{WorldVista}, accounting for 92.86\% and 89.47\% respectively.
While the number of correct identification is 10 and 13, the precision is 76.92\% and 76.47\%.
The reason why there are so many \emph{input-output-interlock} conflicts is that some nouns or noun phrases appear largely in the requirement set.
For instance, the word ``\emph{flight plan}'' appears 21 times, in which 18 times are \texttt{input} and 12 times are \texttt{output} (Note that the input and output may be overlapped).
Therefore, lots of requirements are interdependent caused by the same \texttt{input} or \texttt{output}, meaning that their \texttt{input} and \texttt{output} contain a number of same \texttt{entities}.
As for the false conflicts, one possible reason is that some \texttt{entities} in \texttt{input} or \texttt{output} are recognized incorrectly.
Another reason is not all \emph{input-output-interlock} relationships are real conflicts, some of which are potential risk of inconsistence. 

In addition to \emph{input-output-interlock} relation, other conflict types detected are only \emph{restriction-inconsistency} and \emph{operation-inclusion}, whose identification are all correct.
We think the reason is that requirement inconsistency types are stricter than requirement inclusion and interlock.
While these requirement sets must have been improved several times during the engineering practical and therefore have high qualities, so there is no such conflict in these sets.

For requirement sets with known conflicts, the detected conflicts contain all types of conflicts we defined and are comprehensive.
All 14 known conflicts can be detected, indicating that the our conflict detection rules are well designed and can handle the requirement documents in these two different fields. 
And there is only one false positive conflict, which is \emph{input-output-interlock}, for one \texttt{entity} in \texttt{input} is recognized incorrectly.
If we correct the \texttt{entity}, the false detection can be avoided.
This shows that our conflict detection rules work.

Based on the evaluation on the two kinds of requirement sets, we get the overall measurements by calculating the overall conflict numbers (including the total conflicts, known conflicts, detected conflicts and the correct identified number) and the metrics including the \emph{precision} (i.e., $\frac{Overall correct identified}{Overall detected conflicts}$) and \emph{recall}(i.e., $\frac{Overall correct identified}{Overall known conflicts}$).
For the two requirement sets from academic groups, the \emph{overall precision} is about 78.79\%.
And for the other two sets, the \emph{overall precision} reached 93.33\%, and the total recall rate reached 100.00\%, indicating that it can accurately detect conflicts in the requirement set.

\section{Threats to validity}
\label{sec:threats}

Threats to validity are discussed from internal, external and construct validity.

Internal validity is about how well the experiment is done so that a causal relationship can be concluded from the study.
The biggest threat to internal validity is from manual labeling, which may bring in subjective uncertainty.
To mitigate this threat, the labeling is performed by two authors with two rounds.
Firstly, the first author did the labeling carefully.
Then, the second author joined in and worked with the firstly author to jointly review each of the labeling results, discussed the divergent opinions until agreements were reached on all labels. 

External validity is about the generalizability of our approach.
Our approach includes two parts which are semantic elements identification and conflict detection.
We used the requirements of five systems from different domains created by different groups on the evaluation.
To be specific, three sets were employed for the first part, and four for the second one.
Therefore, we mitigate bias to any particular requirement set.
Another possible threat to generalizability is the types of conflicts.
We collected the seven types mainly from academic researches and our project experience.
However, these can be extended with regards of more domain background and project knowledge.
In the future, we plan to work on more conflict types.

Construct validity discusses the extent to which the goal that is designed to measure is accurately measured.
For conflict detection, we only calculated \emph{precision} for the two requirement sets of \emph{UAV} and \emph{WorldVista} because obtaining the full sets of conflicts requires sound domain knowledge and it is hard to make sure that.
So we prepared two more requirement sets \emph{Telecom Management} and \emph{Solar Power Supply} whose conflict sets are known beforehand.
In the end, we can measure both the \emph{recall} and \emph{precision} of our approach for better evaluation.

\section{Conclusion and the future work}
\label{sec:conclusion}

Requirements conflict is highly likely delaying the progress of software development, and is also one important cause of software instability and imperfect functions.
Most of the previous conflict detection approaches requires (semi-)formal formats of the requirements.
However, natural language requirements is the major representation in industry.
Thus, we proposed an approach  \emph{FSARC} to automatically detect the conflicts between natural language requirements, which analyzes the finer semantic elements in NL requirements, and then detects the conflicts beyond the semantic model of these requirements.
We designed two experiments for evaluating the semantic annotations and conflicts detection respectively involving five open requirements sets.
Evaluation results show that our approach is well-behaved.
To be specific, for semantic annotation, with a total 179 requirements from three requirement sets, the total accuracy of automatic annotation reached 79.33\%.
And for conflict detection, with a total of 242 requirements from four requirement sets, the total recall of the algorithm reached 100.00\%, and the precision reached 83.33\%.

In future, we plan to combine the domain model and general dictionary for obtaining more accurate relationships between the phrases to improve the performance of our approach on both semantic annotation and conflict detection.
Besides, we plan to conduct a systematic review on project experts who have deep domain knowledge and at least a few years project experience, with the expectation of obtaining more conflict types as well as their usual detection methods in practice, for improving our approach.

\section{Acknowledgements}
This work is supported by the National Science Foundation of China Grant No.61672078 and No.61732019.

\bibliography{ref}

\end{document}